\documentclass{JHEP3}

\usepackage{amsfonts,amsbsy,amsmath,amssymb}
\usepackage{epsfig}
\usepackage{natbib}

\newcommand{\thetachar}[2]{\vartheta\left[ \begin{array}{c}
          #1 \\ #2
                    \end{array}\right]}
\newcommand{\be}{\begin{equation}}
\newcommand{\ee}{\end{equation}}
\newcommand{\bea}{\begin{eqnarray}}
\newcommand{\eea}{\end{eqnarray}}

\newcommand{\ddelta}{\sigma}
\newcommand{\xxi}{\xi}
\newcommand{\ttau}{\kappa}
\renewcommand{\aa}{\mathbf a }
\makeatother

\title{Expansion for the solutions of the Bogomolny equations on the torus}
\author{Antonio Gonz\'alez-Arroyo and Alberto Ramos\\ Departamento de
  F\'{\i}sica Te\'orica Universidad Aut\'onoma de   Madrid and\\
  Instituto de F\'{\i}sica Te\'orica UAM-CSIC\\ E-mail:
  \email{tony@martin.ft.uam.es}, \email{alberto@martin.ft.uam.es}}  

\keywords{Bogomolny equations, vortices, abelian higgs model, field theories in lower dimensions, solitons monopoles and instantons}
\preprint{\hepth{0404022}; FTUAM-04-04; IFT-UAM/CSIC-04-08}

\maketitle

\abstract{
We show that the solutions of the Bogomolny equations for the Abelian Higgs
model on a two-dimensional torus, can be expanded in powers of a quantity
$\epsilon$ measuring the departure of the area from the critical area. 
This allows a precise determination of the shape of the solutions for 
all magnetic fluxes and arbitrary position of the Higgs field zeroes. The 
expansion is carried out to  51 orders for a couple of representative cases,
including the unit flux case. We analyse the behaviour of the expansion in 
the limit of large areas, in which case the solutions approach those on the
plane. Our results suggest convergence all the way up to infinite area.
}

\begin{document}

\section{Introduction}

Topological defects play an important role in Particle Physics,
Cosmology, and many areas of Condensed Matter Physics, like
superconductivity, superliquid helium, etc.
Minimum energy(action)  configurations carrying topological charges
arise as solutions of non-linear  partial differential equations.
In some situations, these solutions are not explicitly known, and one needs to 
make use of approximate analytical or numerical methods to study 
their structure and properties. 

The Abelian Higgs model is  one of the simplest models to study 
these ideas. It serves as a relativistic field theory extension of the 
Ginzburg-Landau description of a superconductor, in which 
 a scalar field represents  the condensate of Cooper pairs. 
 In addition to superconductivity, modifications of the
 Abelian Higgs model have found applications in other domains of Physics,
 such as Cosmology (see for example
 Ref.~\cite{vilenkin}). On a different level, the Abelian Higgs model
acts  as  a simplified  model in which to explore  ideas and
develop methods useful for  non-abelian gauge theories. It illustrates 
phenomena such as spontaneously broken gauge symmetry,  colour confinement,
the dual Meissner  effect, topological charges, and
others, all of which  play a role in our present understanding of Particle
Physics.

Abrikosov~\cite{abrikosov} realised that superconductors contain string-like 
topological defects or configurations, which represent  magnetic flux tubes. 
A corresponding stable stationary
cylindrically-symmetric configuration of minimum energy per unit length of 
the Abelian Higgs model was discovered by Nielsen and Olesen~\cite{no:vortice}. 
Cutting a slice orthogonal to the direction of the string, we obtain a 
two-dimensional field configuration. Its stability arises from the fact that it
 possesses non-trivial topological charge. 
In this case, the topological charge is just  the magnetic flux going
through the  plane which, for finite energy (per unit length) configurations,
is quantised in multiple units of $2 \pi$. Despite its conceptual simplicity, 
there is no explicit analytical expression for the unit  magnetic-flux 
minimum energy configuration(the Nielsen-Olesen vortex).

 The self-coupling  constant of the scalar field
$\lambda$ determines the ratio of photon to scalar field masses. There is a
critical value of this coupling $\lambda_c$ (equal to one, in our units),
separating the case of type I ($\lambda < \lambda_c$) and type II ($\lambda
> \lambda_c$)
superconductors~\cite{Kramer:1966,Muller:1966}. Similarly to what happens 
with non-abelian gauge  theories in 4 dimensions, 
Bogomolny~\cite{Bogomolny:1975de}
discovered that (for $\lambda \ge \lambda_c$) the 2-dimensional energy of any 
configuration is bounded  by a given constant  times the 
absolute value of the topological charge. 
At the critical  coupling $\lambda_c$,  the bound is saturated by 
the  solutions of a system of first order partial differential equations:
the Bogomolny equations.

Solutions of the Bogomolny equations not only   exist but, 
for  $\lambda = \lambda_c$, exhaust all solutions of the two-dimensional field 
equations (energy extrema)~\cite{Taubes1980cm}. Furthermore, the space of 
solutions  defines a  manifold  of dimension 
$2q$~\cite{Taubes:1980tm}, where $q$ is the number of flux quanta. 
This is again very similar to the situation for 
self-duality equations in non-abelian gauge theories in 4 dimensions (4D). 
Indeed, a connection between both topics is known~\cite{Taubes1980cm}. 
Despite these properties, there are no explicit analytical expressions for 
these solutions.
The unit flux cylindrically-symmetric solution was constructed as a power series
in the radial coordinate by  H. J. de Vega and F. A. Schaposnik \cite{deVega:1976mi}. 
For higher fluxes, a solution is determined uniquely by the location  
of the $q$ zeroes (counted with multiplicity) of  the Higgs field. 
These points can be interpreted as the position in the plane of $q$ 
unit-vortices. Since the energy does not depend on these positions (is given by the 
flux alone), one can think of these unit-vortices as non-interacting (This
is not exactly true when one takes into account the kinetic terms in 4D).
There are corresponding zero-modes associated with this
degeneracy~\cite{Weinberg:er}.  This contrasts with the situation for
non-critical coupling. 
Numerical simulations \cite{rb:vortice} and other methods
tell us that when $\lambda > \lambda_c$, the interaction between vortices is
repulsive, and there are no static solutions of the equations of motion for
flux greater that $2\pi$  in the plane. On the other hand, if
$\lambda < \lambda_c$ the interaction between vortices is attractive,
and the solution of the equations of motion is a single vortex that
carries all the flux (also called a ``giant'' vortex).

Certain problems require some knowledge of the structure of the solutions,
and, in the absence of analytical expressions for them, one must 
rely upon  approximate methods. Frequently these methods are specific 
to cylindrically symmetric solutions, or make use of 
specific ansatze. This is the case of the numerical methods of 
Ref.~\cite{rb:vortice}, for example. Other approximations are based on 
expansions in powers of certain quantities. We already cited the 
study of de Vega and Schaposnik~\cite{deVega:1976mi} for cylindrically-symmetric 
solutions. Similar expansions  have been used to study excitations of the 
vortices in the type II superconducting
phase~\cite{Arodzycia}.

In this paper we will present a new method to obtain the solutions of
the Bogomolny equations on the 2-dimensional torus by means of an expansion 
in a parameter measuring the departure of the area from the critical area 
of $4q \pi$ (measured in units of the square correlation length).  The method  
can be applied to obtain solutions of arbitrary 
flux and with Higgs zeroes at prescribed (but arbitrary)
points. Truncating this expansion at a given order  
one obtains approximations to the solutions which tend to become worse with
increasing area. However, as we will see, accessible truncations of the
expansion give precise results even for large sizes, where the torus 
solutions approach those of the plane. Thus, although our primary goal is
that of obtaining the solutions on the torus, it seems possible  that 
problems pertaining to the plane can be addressed in this way too. 
Part of our motivation
arises from the fact that a similar expansion was proposed  to study 
self-dual Yang-Mills field configurations on the four dimensional
torus~\cite{GarciaPerez:2000yt}. The present case is a simplified version of 
the 4D problem, and  hence better suited for performing a more  
detailed convergence analysis.

The outline of the paper is as follows. In the next section the method 
will be presented. In the following one,  we will apply it to the two 
simplest cases: The unit-flux vortex case and a flux=2 case with no 
circular symmetry.  In both cases the expansion 
is carried out up to 51 orders. This enables us to do an analysis of 
convergence of the series for various sizes, including the infinite 
area case. Finally, in the last section, we present the conclusions
and indicate prospects for future research. In addition, we have included 
an appendix describing  a variant method with a quantum mechanical 
flavour. Although, less efficient than the method explained in section 
2, it has several  advantages  over it, one  being that, at the present
stage, seems better suited for generalisation to the non-abelian 
self-duality study. It also has served as a test of the actual computed
coefficients  of the expansion, since they have been obtained with both
methods and  two independent codes. 

\section{Description of the method}
\label{sc:method}

In suitable units (unit charge and photon mass), the Lagrangian density 
of the abelian Higgs model in four-dimensional($4D$) 
Minkowski space-time, is given by:
\begin{equation}
  \mathcal L =
  -\frac{1}{4}F_{\mu\nu}F^{\mu\nu} +
  \frac{1}{2}(D_\mu\phi)^*(D^\mu\phi) 
  - \frac{\lambda}{8}(|\phi|^2-1)^2
\end{equation}
where $\phi$ is a complex scalar field, and $D_\mu = \partial_\mu -
iA_\mu$ is the covariant derivative with respect to $A_\mu$, the $U(1)$ gauge
potential. This system is
known~\cite{abrikosov,no:vortice,Taubes:1980tm} to possess  static,
$z$-independent (vortex like)  solutions of the classical equations 
of motion.  These configurations are local minima of  the  $2D$ energy,
whose density is  
\begin{equation}
  \mathcal E = 
  \frac{1}{4}F_{ij}F^{ij} +
  \frac{1}{2}|D_i 
  \phi|^2 + \frac{\lambda}{8}(|\phi|^2-1)^2
\end{equation}
where Latin indices run over two spatial coordinates: $i,j,\dots= 1,
2$. 

We will focus  in the case in which  the  coupling $\lambda$ takes the
critical value $\lambda_c = 1$.  Then, the second order (2D) differential equations
of motion reduce to a set of first order ones (Bogomolny
equations)\cite{Bogomolny:1975de}.
For positive  flux and in our units these equations take the form:
\begin{eqnarray}
\label{firstequation}
  (D_1+iD_2)\phi = 0 \\
\label{eq:bog2}
  B = \frac{1}{2}(1-|\phi|^2)
\end{eqnarray}
where $B$ is the (z-component of the) magnetic field.

Our goal is to study  the solution to these Bogomolny equations for fields 
living in a $2D$ Torus (for an introduction to gauge fields on the torus
see\cite{ga:torus}). The appropriate mathematical description involves
sections of a  U(1) bundle.  We will be working   within a fixed trivialisation. 
The torus can be viewed as the quotient space of the plane 
${\mathbb R}^2$ modulo the lattice $\Lambda$  
generated by two linearly independent 2D vectors $e^{(1)}$,  $e^{(2)}$. 
We assume that the torus is equipped with a flat Riemannian metric, which we
will fix to be Euclidean. 
Within a specific trivialisation, the charged fields(sections)  $\phi(x)\equiv\phi(x_1,x_2)$
are given by
complex valued functions satisfying certain periodicity properties:
\be
\label{BCondA}
\phi(x+e^{(i)})=\Omega_i(x)\,\phi(x)
\ee
The $U(1)$ fields $\Omega_i(x)$ are the transition functions, which have to
satisfy the following consistency conditions
\be
\label{consistency}
\Omega_1(x+e^{(2)})\, \Omega_2(x)= \Omega_2(x+e^{(1)})\, \Omega_1(x)
\ee
The topological properties of the bundle, encoded in the transition functions,
are associated with the first Chern class  $c_1$. Its corresponding integer 
Chern number is known, and will be shown later, to physically correspond to the 
magnetic flux going through the torus in units of $2\pi$. 

Without loss of generality we can choose the following specific form of the 
transition functions: 
\be
\label{BConds}
\Omega_i(x)=\exp\{\imath \pi  \omega(e^{(i)},x)\} 
\ee
where $\omega$ is an antisymmetric form. 
The consistency
condition~(\ref{consistency}) forces \\
$\omega(e^{(1)},e^{(2)})\equiv q$ to take
integer values. This is precisely the first Chern number mentioned 
previously. 

Gauge fields are connections on this bundle. It is well known that the space 
of gauge fields is an affine space. The associated vector space
is the space of 1-forms on the torus. Thus, we can decompose
\be
A=A^{(0)}+\ddelta
\ee
where $\ddelta$ is a 1-form and $A^{(0)}$ is a specific connection. 
For the latter, it is natural to select  a gauge field having constant
field strength $f=F_{1 2}=-F_{2 1}$:
\be
A^{(0)}_i(x)= -\frac{1}{2} F_{i j}x_j 
\ee
Compatibility with the boundary conditions relates the antisymmetric 
matrix $F$ with $\omega$ as follows
\be
 F=2 \pi \omega
\ee
which shows the relation between flux and Chern number. For the 1-form 
 $\ddelta$ we might use Hodge theorem and split it  as a sum of
of an exact, co-exact and a harmonic form:
\be
 A= \pi \omega(x,dx)+v + dg-\delta h
\ee
In components we might write 
\be
A_i(x)= A_i^{(0)}(x) + \partial_i g -\epsilon_{i j} \partial_j h +v_i
\ee
where $h$ and $g$ are real periodic functions and  $v_i$ are constants (a
representative of the harmonic forms on the torus).
Although locally these constants are pure gauges, globally they are not. 
Their value influences  Polyakov loops winding around the torus, which are gauge 
invariant quantities. In this way it  is easy to realise  
that the $v_i$ can be considered elements of the dual torus 
${\mathbb R}^2/\Lambda^*$. In addition, one can fix $h$ so that its 
integral over the torus vanishes ($  \int d^2x\, h=0$).

Now we can return to the Bogomolny equations and express them 
in terms of $v_i$ and the periodic function $h$. For simplicity we will restrict 
to orthogonal periods $e^{(1)}=(l_1,0)$, $e^{(2)}=(0,l_2)$ 
(in appendix~\ref{sc:quantum} we will study the general case).
It is convenient  to make use of   complex coordinates:
\begin{subequations}
  \begin{equation}
    z = \frac{x_1+ix_2}{2};\qquad
    \overline z = \frac{x_1-ix_2}{2}
  \end{equation}
  \begin{equation}
    \partial = \partial_1 - i\partial_2;\qquad 
    \overline\partial=\partial_1+i\partial_2
  \end{equation}
  \begin{equation}
    \partial_1=\frac{1}{2}(\partial+\overline\partial);\qquad
    \partial_2=\frac{1}{2i}(\partial-\overline\partial)
  \end{equation}
\end{subequations}
The notation is chosen so that $\partial z =
\overline\partial\overline z=1$. 

Similarly the vector potential can be 
expressed as a complex function $A = A_1-iA_2$ (with $\overline A =
A_1+iA_2$).  In this notation  the Bogomolny equations take the form:
\begin{subequations}
  \begin{equation}
    (\overline\partial - i\overline A)\phi = 0
  \end{equation}
  \begin{equation}
    -\frac{i}{2}\left(\partial \overline A - \overline\partial A\right) = \frac{1}{2}(1-|\phi|^2)
  \end{equation}
\end{subequations}
Our specific parametrisation of the vector potential 
($A^{(0)} = -if\overline z$) becomes: 
\begin{equation}\label{eq:hodgec}
  A = \partial g - i\partial h + v - if\overline z
\end{equation}
where  $h$ and $g$ are periodic functions, and $v=v_1-iv_2$ is a
complex constant. For the Higgs field we will use  the following
parametrisation 
\begin{equation}
  \label{eq:higgscomp}
\phi =  {\mathcal N} e^{-h+ig}\chi
\end{equation}
where $\chi$ is a normalised function (i.e. $\int_{\mathbb T^2} |\chi|^2
d^2x = l_1l_2$), satisfying  the same boundary conditions as $\phi$,
and $\mathcal N$ is a normalisation constant.
With this terminology the Bogomolny equations become:
\begin{subequations}
 \label{BOGBOTH}
  \begin{equation}
   \label{BOGONE}
    (\overline\partial +f z)\chi = i \bar{v} \chi
  \end{equation}
  \begin{equation}
  \label{BOGTWO}
    \partial\overline\partial h = \frac{1}{2}(1-2f - |{\mathcal
    N}|^2e^{-2h}|\chi|^2) 
  \end{equation}
\end{subequations}
These equations can be solved sequentially. The first equation allows 
one to determine $\chi$, satisfying the correct boundary and normalisation 
conditions. This can be done analytically as will be shown in the next 
subsection. Once this is solved,  we can use the second equation to solve 
for the periodic function $h$ and the normalisation  $\mathcal N$.  
The equation is, however, a non-linear partial differential equation and
the analytic solution is not known. For a particular value of the 
area $l_1l_2=4q\pi$, corresponding to $f=\frac{1}{2}$, a solution is given 
by $h=\mathcal N =0$. Our strategy  consists in using $\epsilon=1-2f$ as a
perturbative parameter.  In what follows we will explain more precisely 
the procedure that we follow.

Since $h$ is periodic, we can use its  Fourier decomposition:
\begin{equation}
  h = \sum_{n_1n_2}h_{n_1n_2}e^{2\pi in_1 x_1/l_1}e^{2\pi in_1 x_2/l_2}
\end{equation}
and expand the Fourier coefficients in a power series in $\epsilon$:
\begin{equation}
  h_{n_1n_2} = \sum_{k=1}^\infty h_{n_1n_2}^{(k)} \epsilon^k
\end{equation}
Similarly, we can write 
\begin{equation}
  |{\mathcal N}|^2 = \sum_{k=1}^\infty \mathtt A_k \epsilon^k
\end{equation}
The only additional input that  we will need are the Fourier coefficients of
$|\chi|^2$. Although in some cases it is interesting and possible to deal with Fourier 
coefficients which are series expansions  in powers $\epsilon$, we will restrict 
here to the case in which these coefficients are independent of $\epsilon$ 
(but  dependent   on the aspect ratio $\tau=\frac{l_2}{l_1}$). The different 
situations will be clarified in the following subsection. 
Notice that our condition on $h$ fixes $h_{0 0}=0$. Alternatively, we might 
have taken $\mathcal N=1$, and allow for non-zero  $h_{0 0}$, but  
our choice is more natural within our expansion.

To solve Eq.~\ref{BOGTWO} we  equate the Fourier coefficients 
of both sides of the equation order by order in $\epsilon$. The left-hand side
takes the form
\be
-(1-\epsilon) \xxi(\tau,n_1,n_2) \sum_k h^{(k)}_{n_1n_2} \epsilon^k
\ee
where 
\be
\label{eq:defxi}
\xxi(\tau,n_1,n_2)=\frac{\pi \tau}{q} \left[
    n_1^2+\frac{n_2^2}{\tau^2} \right]
\ee
which vanishes for $n_1=n_2=0$. We remind the reader that $\tau=l_2/l_1$ and 
$q$ is the first Chern number. 

To treat  the  right-hand side  we first expand it in powers of $\epsilon$. 
To order $\epsilon$ the coefficient is given by $(1-\mathtt A_1|\chi|^2)/2$. 
The coefficient of order $\epsilon^N$ (for $N>1$)  is given by
\begin{equation}
  \label{eq:oN}
  - \frac{1}{2}\sum_{k=0}^{N-1}\mathtt A_{N-k}
  \left[\frac{(-2)^k}{k!}
    \sum_{
      \begin{array}{c}
        i_1,{\ldots} ,i_{N-1} \\
        / \sum_s s i_s = k
      \end{array}
    }\binom{\sum_s i_s}{i_1,{\ldots} ,i_{N-1}}
    {h^{(1)}}^{i_1}\cdots {h^{(N-1)}}^{i_{N-1}}\right]|\chi|^2
\end{equation}

 Now, using the Fourier coefficients of $|\chi|^2$ (see the following subsection), 
 we can  obtain the Fourier coefficients of the expression above by applying
a series of  convolutions. We will label  these Fourier modes with the symbol
$\mathcal F_{n_1n_2}^{(N)}$. These are functions of $\mathtt A_i  (i=1,\dots,N)$, and
$h^{(i)} (i=1,\dots,N-1)$. 
In order for the equation to have a solution it is 
necessary that $\mathcal F_{0 0}^{(N)}=0$. This can be regarded as an equation for
$\mathtt A_N$. It allows to determine this coefficient uniquely in terms of 
$\mathtt A_i  (i=1,\dots,N-1)$, and
$h^{(i)}  (i=1,\dots,N-1)$. 

Finally, the equation allows one  to  obtain the Fourier coefficients
$h^{(N)}_{n_1n_2}$ to order $\epsilon^N$ as follows:
\begin{equation}
  \label{eq:rec}
  h_{n_1n_2}^{(N)} = \left\{ \begin{array}{ll}
    -\frac{\mathcal F_{n_1n_2}^{(N)}}{\xxi(\tau,n_1,n_2)} +
h_{n_1n_2}^{(N-1)} & n_1\ne 0 \quad \mbox{or}  \quad  n_2\ne 0 \\
      0 & n_1=n_2=0\\
    \end{array}\right.
\end{equation}
This equation defines a recurrence which, starting from  $h^{(0)}_{n_1n_2}=0$, 
 enables the determination of  the coefficients  $h^{(k)}_{n_1n_2}$ uniquely. 
To order $\epsilon$,  $\mathcal F_{n_1n_2}^{(1)}$ are the Fourier coefficients of 
$(1-\mathtt A_1|\chi|^2)$. The vanishing of the $n_1=n_2=0$ coefficient and the 
normalisation condition for $\chi$ implies $A_1$=1. The coefficients 
$h_{n_1n_2}^{(1)}$  are then given by
\be
h_{n_1n_2}^{(1)} = -(1-\delta_{n_1 0}\delta_{n_2 0}) \frac{\mathcal
      F_{n_1n_2}^{(1)}}{ \xxi(\tau,n_1,n_2)} 
\end{equation}
\newline

Computing the  coefficients to higher orders demands performing convolutions, 
which involve infinite sums over several integers. We do not have closed analytical
expressions for them. In practice, however, the Fourier coefficients 
decrease very fast with the order, so that a truncation of these sums to a
finite subset allows, as we will show later,  the numerical determination 
of the coefficients to machine precision up to  high orders in the expansion. 

We emphasise that the previous procedure gives rise to a unique solution 
$h$. All the  degrees of freedom associated to the
space of solutions of the Bogomolny equations resides in the function $\chi$
which will be treated in the following subsection.

\subsection{The function $\chi$}

In this section we will solve the first Bogomolny equation, Eq.~\ref{BOGONE}.
First of all we will analyse the dependence on $\bar{v}=v_1+\imath v_2$.
It is easy to see that if $\chi_{\bar{v}}$ is the solution  the equation for a
value of $\bar{v}$, then a solution for $\bar{v}' = \bar{v} +2if a$ is given
by
\begin{equation}
  \chi_{\bar{v}'} = e^{f(\overline a z - a\overline
  z)}\chi_{\bar{v}}(z+a,\overline z+\overline a) 
\end{equation}
Since the prefactor is simply a phase, it is clear that solutions 
corresponding to different values of $v$ represent solutions 
translated in space. Having this point in mind we can simply 
restrict from now on to $v=0$.

To solve  Eq.~\ref{BOGONE} we define  $\eta = e^{f  z(\overline z - z)}\chi$.
In terms of this function the first Bogomolny equation is   
equivalent to the condition of holomorphicity $\overline \partial \eta = 0$. 
So $\eta = \eta(z)$ is
an analytic function of the variable $z$. The boundary conditions for
this function can be obtained from the ones of $\phi$
(Eq.~\ref{BCondA}), and are given by 
\begin{subequations}
  \begin{equation}
    \eta\left(z+\frac{l_1}{2}\right) = \eta(z)
  \end{equation}
  \begin{equation}
    \eta\left(z+i\frac{l_2}{2}\right) = e^{-2if l_2 z +
    f\frac{l_2^2}{2}}\eta(z) 
  \end{equation}
\end{subequations}
These are the typical conditions satisfied by  theta functions~\cite{ww:analysis}.
We will now analyse the space of solutions in various cases.

For minimal flux $q=1$ $f = \frac{2\pi}{l_1l_2}$, it is easy to
see that our function is given by
\begin{equation}
  \eta =\left(\frac{2l_2}{l_1}\right)^{\frac{1}{4}} \vartheta_3\left(\frac {2\pi 
      z}{l_1}|i\frac{l_2}{l_1}\right)  
\end{equation}
This, together with the normalisation condition, allows us to obtain 
the Fourier decomposition of $\chi^2$:
\begin{equation}
  |\chi|^2 = \sum_{n_1,n_2} 
  e^{-\frac{\xxi(\tau,n_1,n_2)}{2}}
  e^{2\pi in_1x_1/l_1}e^{2\pi in_2x_2/l_2} (-1)^{n_1n_2}
\end{equation}
where $\xxi(\tau,n_1,n_2)$ is given by Eq.~\ref{eq:defxi}.
From here it is trivial to obtain  $\mathcal F_{n_1n_2}^{(1)}$ to initiate 
the iteration that determines $h$. 

For flux $2\pi q$ the space of solutions is multiple dimensional. 
There are several alternative ways to characterise individual solutions.
One possibility is to fix the position of the zeroes of $\chi$, which 
coincide with those of the Higgs field $\phi$. Then we can write 
\begin{equation}
  \label{eq:arb1}
  \eta \propto \vartheta_3\left(\frac {2\pi
      (z+z_c-\omega_1)}{l_1}|i\frac{l_2}{l_1}\right)
  \vartheta_3\left(\frac {2\pi
      (z+z_c-\omega_2)}{l_1}|i\frac{l_2}{l_1}\right)
  \cdots
  \vartheta_3\left(\frac {2\pi
      (z+z_c-\omega_q)}{l_1}|i\frac{l_2}{l_1}\right)
\end{equation}
where $\omega_i$ are complex constants, and $z_c=\frac{1}{4}(l_1+il_2)$. 
This is a holomorphic function,
and satisfies the  correct boundary conditions provided
\begin{equation}
  \label{eq:condition}
  \sum_i \omega_i = qz_c
\end{equation}
The function $\eta$ defined in this way has $q$ zeros located at the
points $\omega_i$.
Eq.~\ref{eq:condition} then specifies that the centre of mass of all the
zeroes is located at the centre of the  torus ($\frac{l_1}{2},\frac{l_2}{2}$).
To shift the centre of mass one must choose $v\ne 0$.

An alternative description of the space of solutions  follows naturally
from the quantum mechanical formulation of appendix~\ref{sc:quantum}. 
A basis of the space of holomorphic functions satisfying the same  
boundary conditions as  $\eta(z)$ is  given by:
\begin{equation}
  \eta_s(z)=\thetachar{s/q}{0}\left(\frac{2\pi q z}{l_1}|iq\frac{l_2}{l_1}
  \right)\qquad (s=0,\dots,q-1)
\end{equation}
where the symbols $\thetachar{a}{b}$ denote Theta functions with 
rational characteristics:
\begin{equation}
  \vartheta\left[ \begin{array}{c} 
      a \\ b 
    \end{array} \right](z | \tau) = 
  \sum_{n\in\mathbb Z} e^{i\pi\tau (n+a)^2}e^{2i (n+a)(z+b)}
\end{equation}
The functions $\eta_s(z)$ have $q$ zeros located at
\begin{equation}
  \left(\frac{l_1}{2q}(2k+1), \frac{l_2}{2}-\frac{s}{q}l_2  \right)
  \qquad (k=0,\dots,q-1)
\end{equation}
An arbitrary solution of the problem is given by a suitably normalised linear 
combination of $\eta_s(z)$: $\eta\propto\sum_s c_s \eta_s(z)$.

Now we can relate the two descriptions by relating the coefficients 
$c_s$ to the position of the zeroes $\omega_i$. 
We can construct a linear combination that possesses  $q-1$ zeros at the
(distinct) complex points $\omega_i$, $i=1,\dots,q-1$ by setting 
\begin{equation}
  \label{eq:arb2}
  \eta(z) \propto  \left|
  \begin{array}{cccc}
    \eta_0(z) & \eta_1(z) & \cdots & \eta_{q-1}(z) \\
    \eta_0(\omega_1) & \eta_1(\omega_1) & \cdots & \eta_{q-1}(\omega_1) \\
    \vdots & \vdots & \vdots & \vdots \\
    \eta_0(\omega_{q-1}) & \eta_1(\omega_{q-1}) & \cdots & \eta_{q-1}(\omega_{q-1}) \\
  \end{array}\right|
\end{equation}
General theorems about elliptic functions tell us that the remaining
zero of this $\eta$ function is located at a point  $\omega_0$ 
which enforces the centre-of-mass condition.

The coefficients
\begin{equation}
\label{CTOZERO}
c_s = \varepsilon_{s i_1\dots  i_{q-1}}\eta_{i_1}(\omega_1)\cdots\eta_{i_{q-1}}(\omega_{q-1})
\end{equation}
give rise to coordinates in the manifold  of solutions of the Bogomolny 
equations.
Actually, they define coordinates in the submanifold defined by the centre of mass
condition for the zeroes Eq.~\ref{eq:condition}. Using properties of elliptic functions and the
result of Taubes~\cite{Taubes:1980tm} one can easily see that this submanifold is
diffeomorphic to $\mathbf{CP}^{q-1}$ and $c_s$ are homogeneous coordinates.
One has to include $v$ to give coordinates over the whole manifold of
solutions. Since the latter varies over a torus, we recover the 
fiber bundle structure described in
Refs.~\cite{Shah:1993us,Nasir:1998,Nasir:1999}.

The formula relating the coordinates to the zeroes Eq.~\ref{CTOZERO}
fails if two or more zeroes coincide. To write the correct formula 
one has  to substitute some  rows of Eq.~\ref{eq:arb2} with the
values of the derivatives (up to  order given by the degeneracy) 
of the functions $\eta_s$. 

Finally,  it is easy to obtain 
the Fourier coefficients of  $|\chi|^2$, using the corresponding ones
for theta functions with rational characteristics. The result is 
\begin{equation}
  |\chi|^2 = \frac{1}{\sum_s |c_s|^2} \sum_{n_1,n_2} 
  e^{-\frac{\xxi(\tau,n_1,n_2)}{2}}
  e^{i\pi\frac{n_1n_2}{q}}\left( \sum_s c_sc^*_{s+n_1} e^{2\pi i\frac{sn_2}{q}} \right)
  e^{2\pi in_1\frac{x_1}{l_1}}e^{2\pi in_2\frac{x_2}{l_2}}
\end{equation}
where $c_s$ is given above.

In all our previous construction the position of the zeroes of the 
Higgs field have been fixed \emph{relative} to the Torus lengths $l_1,l_2$.
 One could, in principle,  insist in  fixing  the position of the zeroes 
 in units of the inverse photon mass.  This amounts to taking the positions
$\omega_i$ as functions of $\epsilon$.
 We might still use the previous formula but now one  has 
 \begin{equation}
   |\chi|^2 = \sum_k C_k(x)\epsilon^k
\end{equation}
The iterative solution of Eq.~\ref{BOGTWO} given in the previous subsection
must then be modified by combining the  powers of $\epsilon$ from $ |\chi|^2$ 
into the expansion.

\section{Representative  examples}

In this section we will show the effectiveness of the method 
presented in the previous section, by applying it to two explicit 
examples. The first one is the standard $q=1$ case, which  \emph{should} 
converge towards  the usual Nielsen-Olesen vortex with cylindrical 
symmetry when the Torus is big enough. An extensive analysis of convergence 
of the series for different volumes will be presented.
The second example is a $q=2$ 
case which has no remnant continuous spatial symmetry. This is important since 
many alternative methods are specific to cylindrical symmetry.

\subsection{$q=1$}

We have applied the machinery explained in the previous section to 
the unit flux $q=1$  and unit aspect ratio $\tau=l_2/l_1=1$ case. 
The solution is unique up to translations.
We have obtained the coefficients of the expansion $h_{n_1n_2}^{(k)}$
up to order $k=51$. Convolutions were performed by truncating 
the sums to Fourier modes in the range $|n_i|\le n_{\rm \tiny max}$. 
This was a implemented  using a \texttt{FORTRAN 90}  program running
in a standard PC. Results up to  $k=51$ required computation times
around 100 hours. 

We have, first of all, analysed the effect of truncation in the number 
$n_{\rm \tiny max}$ of Fourier modes used in convolutions on the value 
of the coefficients. We explored the cases  $n_{\rm \tiny
  max}=10,14$ and $20$. 
We estimate the relative difference (difference over sum)
in the coefficient $h_{n_1n_2}^{(k)}$ obtained from two different 
truncations 10-14 or 14-20. The difference is attributed to an error in 
the value for the  smaller $n_{\rm \tiny max}$. The differences increase 
with $n_i$ and with $k$. However, even for the highest order ($k=51$) the 
relative difference between the results with $n_{\rm \tiny max}=14$ and $20$ is
of the order of machine precision $10^{-16,-17}$ for all 
modes having $n_i < 7$. For higher modes it is to be expected that the 
error is mostly due to an error in the values for $n_{\rm \tiny max}=14$,
so the modes for $n_{\rm \tiny max}=20$ remain probably within machine precision
for higher $n_i$. 
To analyse this we used the comparison 10-14 to try to understand the
dependence of errors with $n_i$, $k$ and $n_{\rm \tiny max}$. It is 
to be expected that the error is proportional to (some power) of the size
of the neglected terms $h_{n_{\rm \tiny max} n_{\rm \tiny max}}^{(k)}$. 
This fits nicely with the observed dependence of the relative difference
\begin{equation}
 -3(1)-\frac{1270(30)}{k} + \max(n_1,n_2)\left(-0.15(6)+ \frac{54(1)}{k}\right)
\end{equation}
Our interpretation of the source of the errors suggest that the 
coefficients multiplying  $\frac{1}{k}$ are proportional to
$n^2_{\rm \tiny max}$. This allows us to scale the results to 
$n_{\rm \tiny max}=14$. Indeed, our data on the difference 
14-20 is consistent with this interpretation, although there are 
few values of $k$ and $n_i$ to allow an analysis by itself. 
On the basis of these results, we conclude that, even at $k=51$, the
coefficients obtained for $n_{\rm \tiny max}=20$ are correct up to
machine (double) precision for $n_i\le 14-15$.

Once the coefficients are obtained, one can reconstruct the Fourier 
modes of $h$, the magnetic field $B$ and the modulus of the Higgs field
$|\phi|$ for arbitrary torus areas ${\cal A}=l_1l_2= \frac{4 \pi}{1-\epsilon}$.
Obviously, the precision of the truncated series  depends on the value 
of $\epsilon$, decreasing as $\epsilon$ increases up to the infinite area 
case of $\epsilon=1$. Even for relatively large areas the shape of the 
reconstructed function looks qualitatively quite good. See for example 
Fig.~1,2, where we display  the magnetic field for
$\epsilon=0.9$,  obtained from the Fourier modes ($|n_i|\le 20$) 
computed with 51 orders in the expansion.

\DOUBLEFIGURE[h]{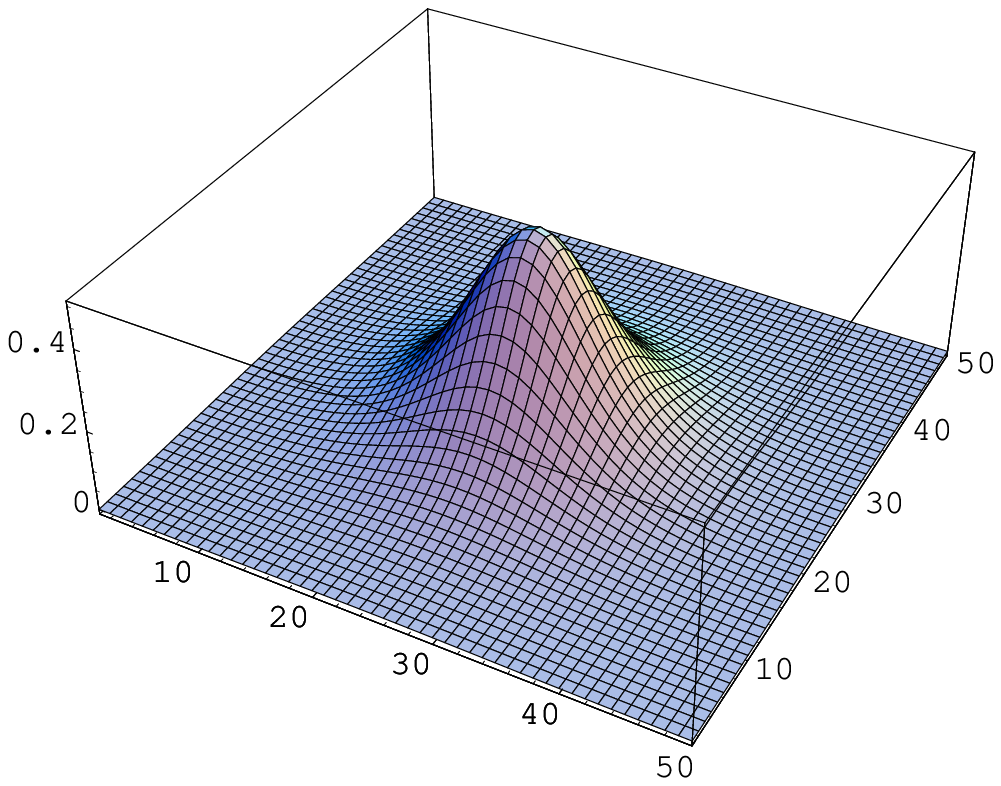,width=6cm}{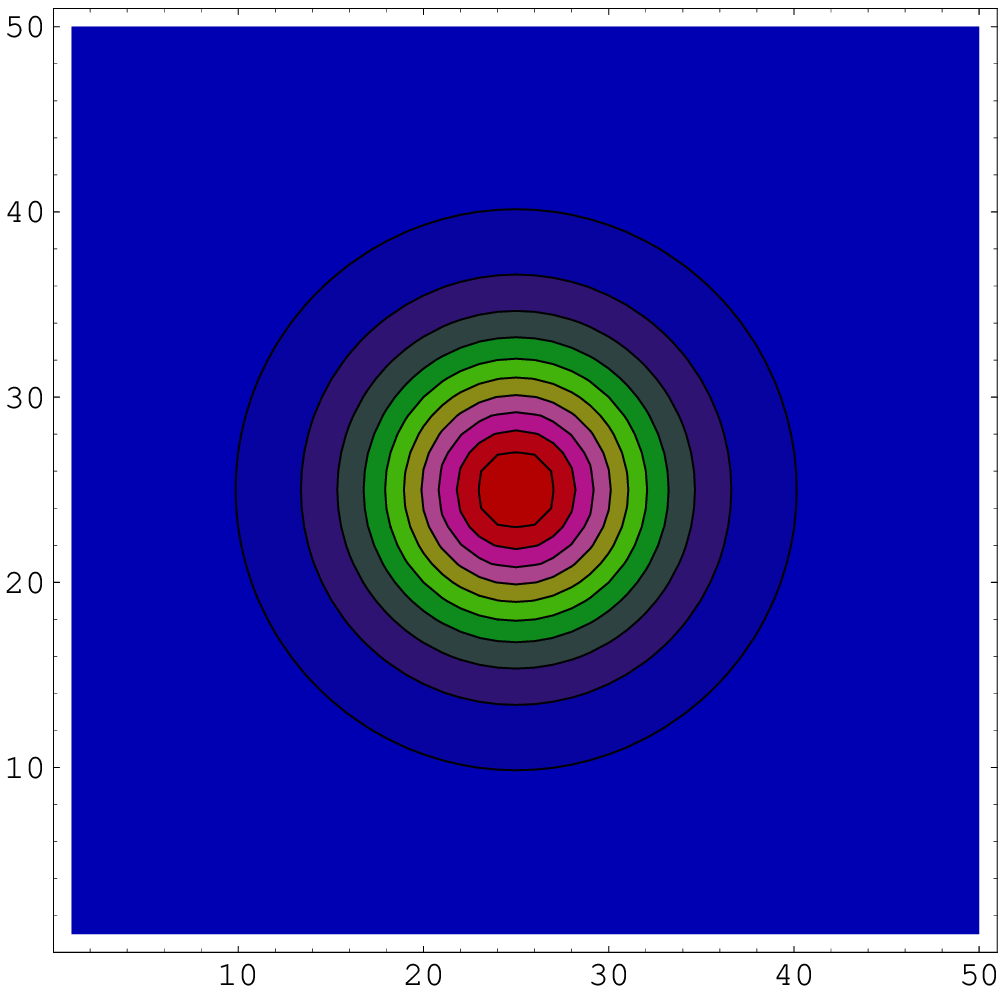,width=6cm}{Shape
  of the magnetic field $B(x)$ obtained for $\epsilon=0.9$ with  51
  orders in the expansion.}{Contour plot of $B(x)$.}

To go beyond the qualitative level and estimate the accuracy of 
the truncated series, we use the degree of satisfaction of the
Bogomolny equation as a measure of the error. Thus, we compute 
the magnetic field $B(x)$ for various values of $\epsilon$ and 
compare it with the right-hand side of Eq.~\ref{eq:bog2} 
\ \ $\frac{1}{2}(1-|\phi(x)|^2)$. The latter is computed using the truncated 
expansion of $h$ and the parametrisation Eq.~\ref{eq:higgscomp}. More
precisely we computed  
the $L_2$ and $L_\infty$ norm of the difference:
\begin{subequations}
  \begin{equation}
    L_2(N,\epsilon) = \left( \frac{1}{l_1 l_2} \int_{\mathbb T^2} dx\, \left[ B(x) -
    \frac{1}{2}(1-|\phi(x)|^2)\right]^2 \right)^{1/2} \quad 
  \end{equation}
  \begin{equation}
    L_{\infty}(N,\epsilon) = \max_x\left\{ \left|B(x) -  \frac{1}{2}(1-|\phi(x)|^2)\right| \right\}
  \end{equation}
\end{subequations}
where $N$ is the maximal order in the expansion. We have
analysed the $N$ and $\epsilon$ dependence of both quantities. Results are
qualitatively the same for both, so we will choose  $L_2$ to display.
First, we will comment on the maximum precision, attained   for $n=51$. 
For $\epsilon\le 0.6$ the $L_2$ norm is compatible with zero within machine 
precision (order $10^{-16,-17}$). Beyond this value $L_2$ becomes sizable and
increases, reaching $10^{-4}$ at $\epsilon=0.95$. For comparison we point out 
that with $N=1$ the value (at $\epsilon=0.95$) is $\mathcal
O(10^{-2})$. Convergence is therefore slow in this case, but notice
that the linear size of the box is 15 times the Debye screening length
or 4.5 times the square root of the critical area.

\DOUBLEFIGURE[h]{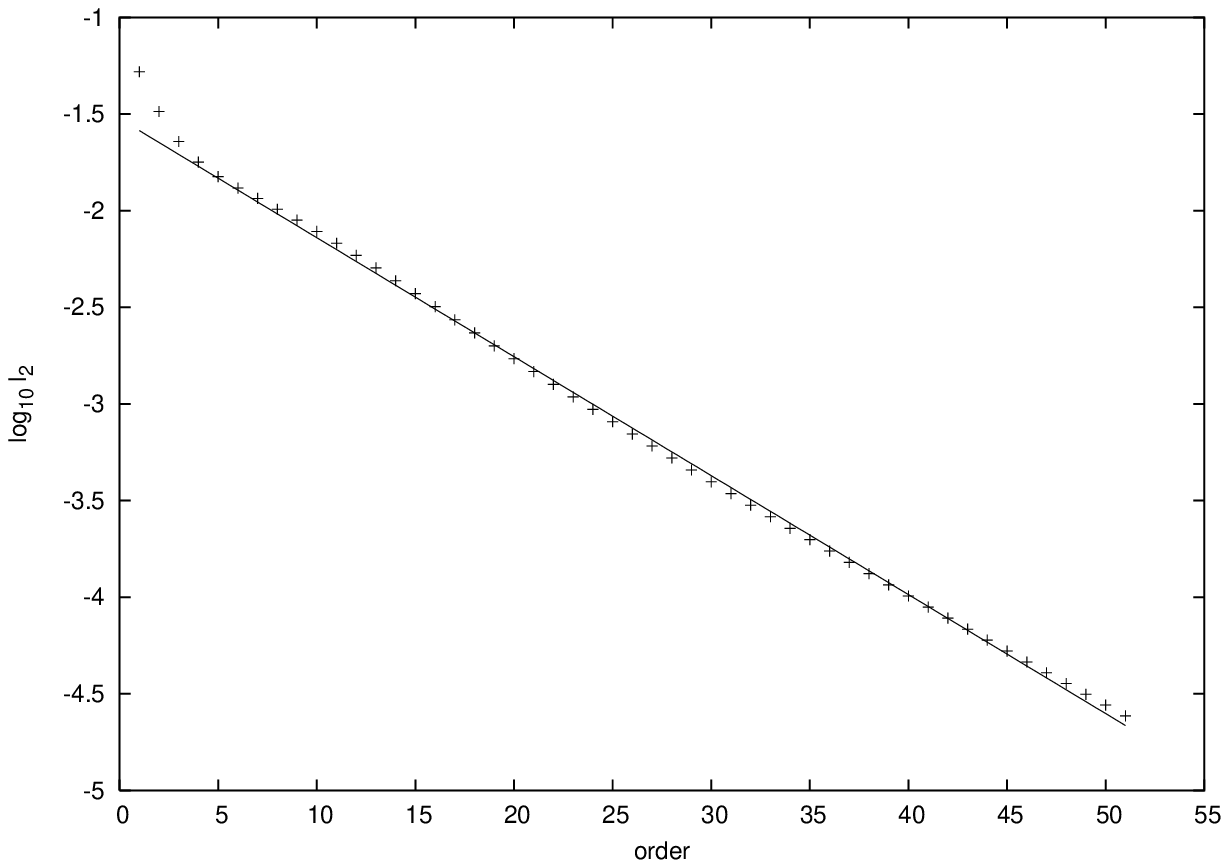,width=6cm}{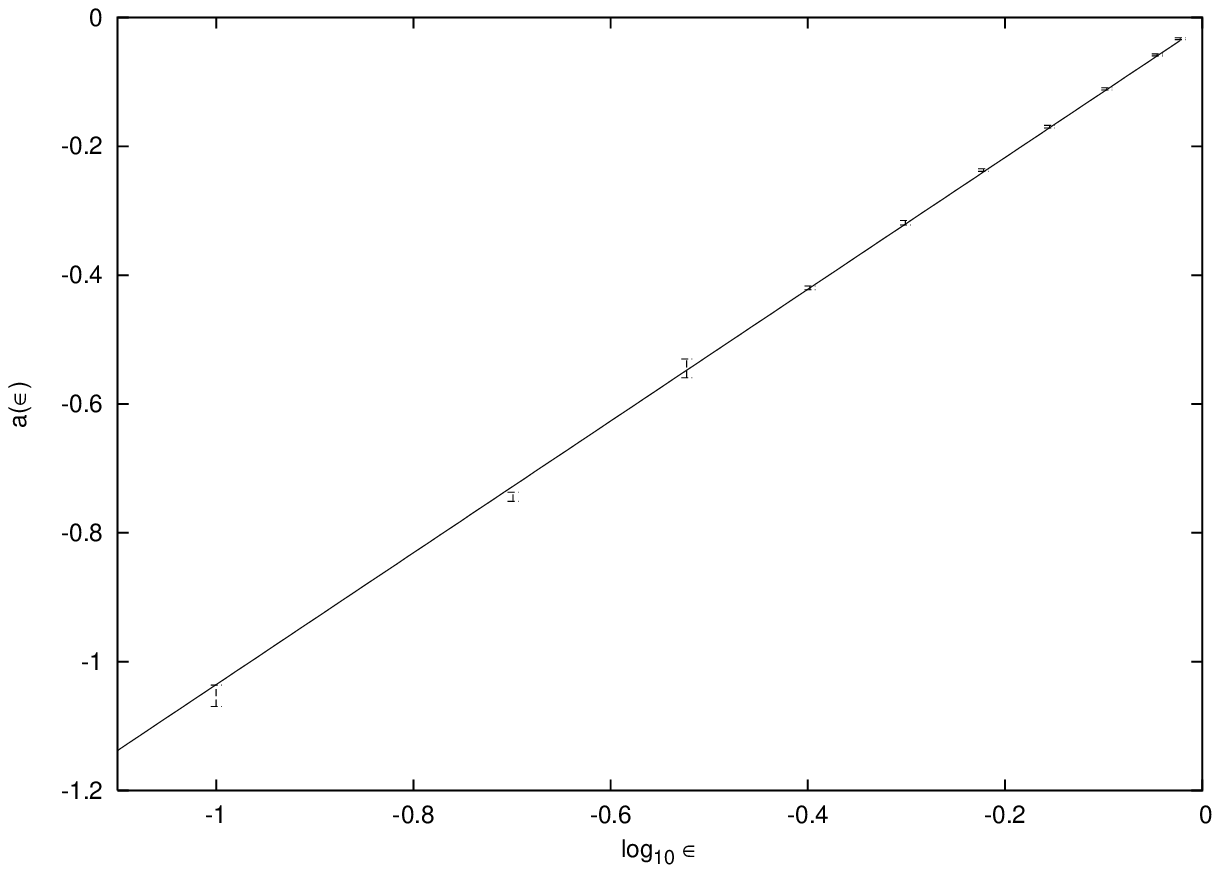,width=6cm}{$L_2(N,\epsilon)$
  versus  $N$ (in $\log_{10}$ scale) for
  $\epsilon=0.9$.}{$a(\epsilon)$ versus  $\epsilon$ compared to our
  best fit Eq.~3.5.}

We performed a more systematic study by analysing the  $N$ dependence for 
fixed value of $\epsilon$, in the range   $0.1-0.95$. In this range the 
results are unaffected by the truncation in the number of Fourier modes 
$n_{\rm \tiny max}$. In all cases, we found that, beyond the first few orders, 
the dependence of $L_2(N,\epsilon)$ with  $N$ oscillates around an
exponential fall-off. As an  example, we show in Fig.~3  
the  $\epsilon=0.9$ case . Therefore, we fitted $\log_{10}L_2$ data to the
following linear function:
\begin{equation}
  \log_{10} L_2(N,\epsilon) = a(\epsilon)N + b(\epsilon)
  \end{equation}
The parameter $a(\epsilon)$ is determined with errors reflecting the
statistical and systematic uncertainties (range of fitting for example). 
Its value determines how the approximation improves when 
increasing the order $N$ in the expansion. Its negative value is an indication
that the expansion is indeed convergent. Obviously as $\epsilon$ increases 
so does $a(\epsilon)$. This is displayed in Fig.~4. For a convergent series 
and small $\epsilon$ one expects 
\begin{equation}
  \log_{10} L_2(N,\epsilon) \approx (N+1)\log_{10}(\epsilon)+
\log_{10}(c_N)
    \end{equation}
    and hence, 
$a(\epsilon)=\log_{10}(\epsilon)+ \mbox{constant}+ O(\epsilon)$. 
This is indeed the behaviour shown by  the Fig.~3.
Fitting  $a(\epsilon)$ to a linear function of 
$\log_{10}\epsilon$ gives:
\begin{equation}
  \label{eq:bestfita}
  a(\epsilon) = 1.022(5)\log_{10}\epsilon - 0.0106(6)
\end{equation}
Errors reflect the quality of the fit. 
Remarkably nothing seems to be happening at $\epsilon=1$, where the area diverges.
Data cannot be taken directly at $\epsilon=1$ because they are severely
affected by the truncation in the number of Fourier modes, but the 
behaviour up to $\epsilon=0.95$ shows no sign of a change of pattern and 
extrapolates to $a(1)<1$. Similar smooth behaviour is shown by $b(\epsilon)$.
So we take our results as an indication that the series actually converges
all the way up to $\epsilon=1$. 

Within our spirit of identifying the $L_2$ (or $L_\infty$) norm of the
equation with the error on the Higgs and magnetic field, we can use our data 
from a more practical viewpoint as an estimate of the number of terms
required in the expansion to attain an a priori decided precision. An
approximate formula can be derived from our data. If one is willing to 
compute the magnetic field with an error of $10^{-p}$ then the number of 
terms required in the expansion is given by:
\begin{equation}
  n \approx \frac{1.83\epsilon + p - 3.25}{0.01 - 1.02\log_{10}\epsilon}
  \end{equation}
Although the formula gives a finite number even for $\epsilon=1$, we
stress once more that in practise at that very large volumes, 
truncation in the number of Fourier of terms would make the expansion
increasingly computationally costly.

Now we will explore the  implications of our expansion for large volumes. 
Our main assumption is that the solutions on the torus do converge to 
those on the plane. The convergence is expected to be fairly fast. 
A  torus configuration is equivalent to a periodic array of vortices on the plane. 
However, vortices are exponentially localised objects so that if the period
is large compared to the typical size of a vortex, the effect of the replicas 
is presumably very small. Now the convergence of the solution implies the following
behaviour of the Fourier modes:
\be
\label{YDEF}
Y(\vec{p})\equiv \frac{\hat{B}(p)}{4\pi q}=\lim_{\epsilon \rightarrow
1}\xxi(1,n_1(p),n_2(p))\, 
h_{n_1(p),n_2(p)}\, (-1)^{n_1(p)+n_2(p)}
\ee
where $\hat{B}(\vec{p})$ is the Fourier transform of the magnetic field on the
plane. The limit is taken at fixed $p$ given by:
\be
p^2\equiv |\vec{p}|^2=\xxi\, (1-\epsilon)
\ee
where  $\xxi$ is given by Eq.~\ref{eq:defxi}. This means that as $\epsilon$ tends to 1,
the integers $n_i(p)$ have to grow. If instead, we take the limit
$\epsilon\rightarrow 1$ keeping $n_i$ fixed, the values should converge to 
$Y(0)=0.5$ irrespective of $n_i$. In our case ($q=1$), computing the value at 
$\epsilon=1$ using our 51 orders and $n_1=1$,$n_2=0$ we get 
$Y(0)=0.499947172199$.
Worse results follow for higher modes ($0.499572$ for $n_1=n_2=1$, 
$0.465354$ for $n_1=2$, $n_2=1$, $0.387814$
 for $n_1=n_2=2$, etc). The numerical agreement provides an additional hint
that the expansion converges up to $\epsilon=1$. It also indicates a 
poorer convergence for larger $n_i$ (see later).

For $p$ non-zero, Eq.~\ref{YDEF} and the expectation of fast
convergence, suggests that tuning $\epsilon$ and $n_i$ in such a way that 
$p$ is fixed we should obtain similar values. Only $p$, the modulus of 
$\vec{p}$, matters due to the cylindrical symmetry of the $q=1$ solution on
the plane. This is also satisfied by our expansion to a fairly high
precision. For example, $Y(p)$ can be computed for $p^2=\pi/20$ using 
$\epsilon=0.975$ and $n_1=n_2=1$, or $\epsilon=0.95$ and $n_1=1, n_2=0$.
From our expansion we get $0.39967$ and $0.39977$ respectively. This 
number is presumably very close ($<1 \%$)  to $Y(p)$ on the plane. 
Similarly for  $p^2=\pi/10$ we get $0.326083$ and $0.326075$ from the same
two modes and $\epsilon=0.95,0.9$ respectively. For $p^2=\pi/5$ we get 
$0.22627$ ($\epsilon=0.8$), $0.22676$ ($\epsilon=0.9$) and  $0.22608$ 
($\epsilon=0.95$).  In this way we can use our expansion to compute  the 
Fourier transform of the magnetic field for a vortex on the plane with a 
precision of a few percent.  

Now we will try to extract the consequences of the  convergence of the 
expansion for the Fourier modes to a universal function of $p$ as 
$\epsilon \rightarrow 1$. For very large $n_i$ (large $\xxi$) 
we can compute  $Y(p)$ by taking $\epsilon= 1-p^2/\xi$. Thus,  
\be
Y(p) \approx -  \xxi\, (-1)^{n_1+n_2}\, h_{n_1n_2}(\epsilon)\approx  -\xxi \sum_{k=0}^{N} 
(-1)^{n_1+n_2}\, h_{n_1 n_2}^{(k)}
e^{-p^2\frac{k}{\xxi}}
\ee
This suggests that for $\xxi\rightarrow \infty$
\be
\label{HCOEFS}
h_{n_1 n_2}^{(k)} \longrightarrow \frac{-1}{\xxi^2}\, \varphi(k/\xxi)\, (-1)^{n_1+n_2}
\ee
where the function $\varphi(y)$ satisfies
\be
Y(p)= \int_0^\infty dy\, \varphi(y)\, e^{-p^2 y} 
\ee
Fourier transforming we get:
\be
\label{BEXP}
B(x)=  \int_0^\infty \frac{dy}{y} \varphi(y)\, e^{-|\vec{x}|^2/(4 y)}
\ee

One can test these considerations by computing approximants to $\varphi(y)$
by using Eq.~\ref{HCOEFS} for finite $n_i$. In Fig.~5 we show the shape 
obtained from the coefficients $h_{n_1n_2}^{(k)}$. For any $y$ we plot only 
those values of $n_i$ such that $\xxi \ge y/25$. The smoothness and small
dispersion of values agrees with our expectations. We also investigated
the way in which the limit is approached for large $\xxi$. For example in 
Fig.~6 we plot $-\xxi^2 (-1)^{n_1+n_2} h_{n_1n_2}^{(k)}$ for 
different values of $n_i$ and a fixed value of $y=k/\xxi=1/\pi$.
The solid line is the result of a fit to a function of the form $a+\frac{b}{\xxi}$. 
Similar behaviour obtains for other $y$ values. This analysis  could be used to 
obtain a more precise estimation of the value $\varphi(y)$. For the time
being we simply used the non-extrapolated shape shown in Fig.~5
and analysed the behaviour for large and small values of the argument $y$.
 For small $y$, the function is well described by $\exp\{a'-b'/y\}$  with 
$a'$ and $b'$ very close  to 1. For large $y$ the behaviour is also very well 
described by an exponential $\exp\{-a-by\}$. A fit in the range $y\in[2 , 6]$
gives  $a=0.2973$ and $b=0.9443$. Assuming our formula  Eq.~\ref{BEXP}, we
can, by saddle point methods,   relate the large $|x|$ behaviour of 
$B(x)$ to these parameters. Indeed, $b$ is predicted to be $1$. The parameter
$a$ is given by $-\log(Z_1/2)$ where $Z_1$ was obtained numerically by 
de Vega and Schaposnik ($Z_1=1.7079$)\cite{deVega:1976mi}, and
recently Ref.~\cite{Tong:2002rq}
predicted its value to be $\log(2)/4$. These values of $Z_1/2$ differ by 10\%
from $e^{-a}$. This is a quite satisfactory agreement for the non-extrapolated 
curve obtained from the coefficients of our expansion.

\DOUBLEFIGURE[h]{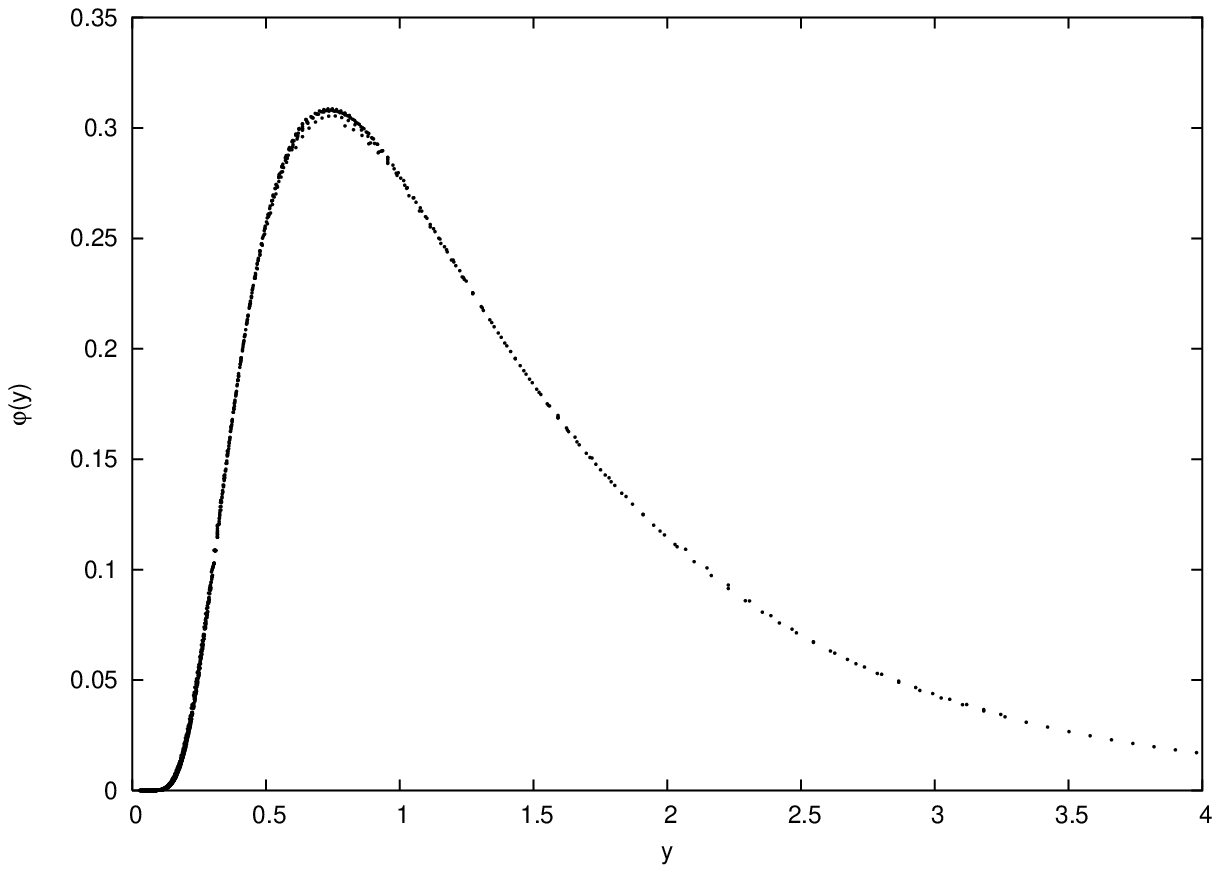,width=6cm}{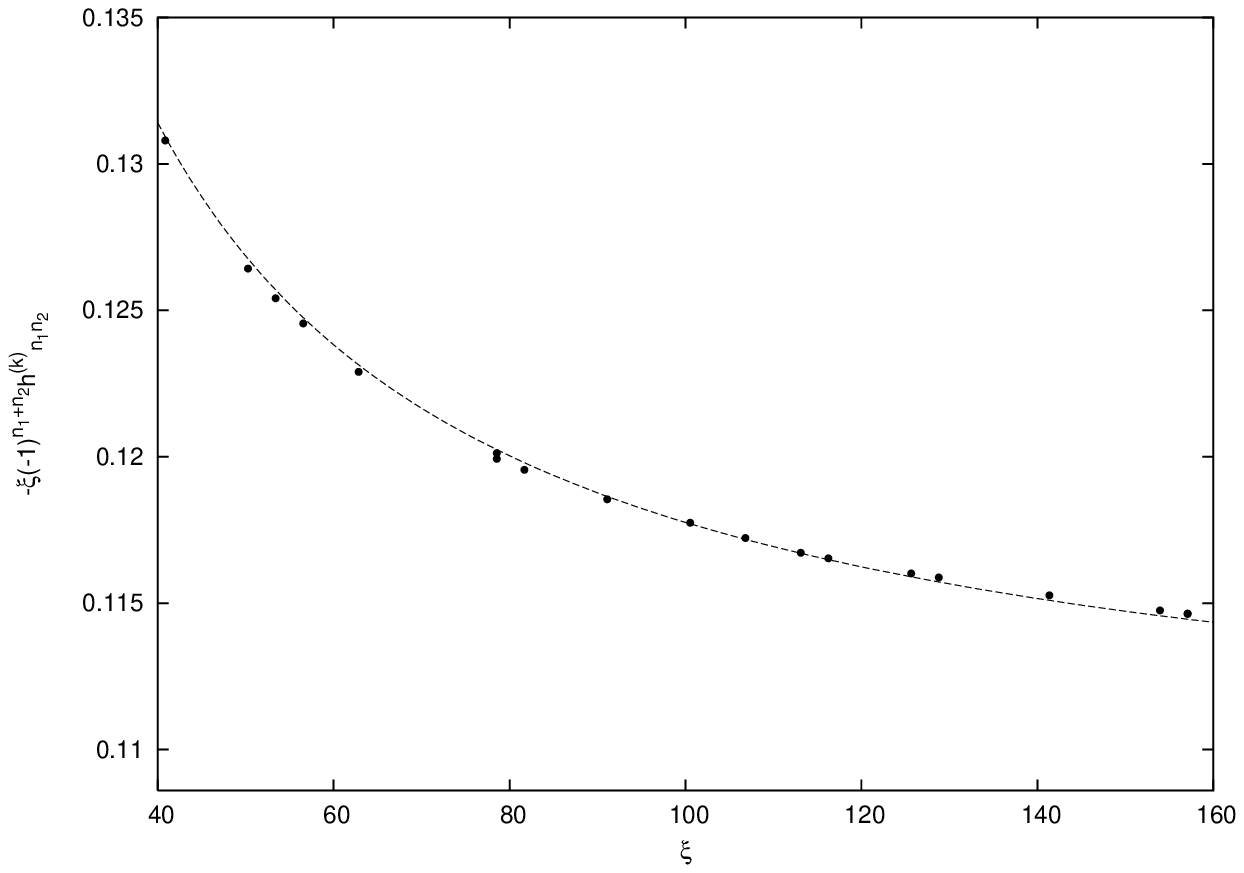,width=6cm}{$\varphi(y)$
  vs. $y$ computed using Eq.~3.10}{$-\xxi^2 (-1)^{n_1+n_2}
  h_{n_1n_2}^{(k)}$ for different values of $\xxi$ and $y=k/\xi=1/\pi$
  fixed.} 

From formula \ref{BEXP} one can deduce a connection between our expansion
and that of Ref.~\cite{deVega:1976mi} in powers of $|x|^2$. The expression
becomes
\be
D_s= 2 \frac{(-1)^s}{s!} \int_0^\infty dz\, z^{s-2}\, \varphi(\frac{1}{z})
\ee
Numerically integrating the data we get $D_1=-0.999976$, 
$D_2=0.747034$, $D_3=-0.523573$, $D_4 = 0.36505$,
in good agreement with Ref.\cite{deVega:1976mi} ($-1$, $0.72791$,
$-0.48527$, $0.31444$ respectively).

From all the discussion above we see that our expansion, though originating 
from a small volume expansion on the torus, matches nicely with results known 
for the single vortex case at infinite area. Even though it might not give
the same level of precision as other methods in that regime, it has the 
advantage of being readily generalisable to  arbitrary fluxes and positions
of the Higgs field zeroes. Furthermore, the same coefficients provide solutions in
arbitrary torus sizes. 

We can actually employ the previous formulas to 
estimate the error committed in $h_{n_1 n_2}(\epsilon)$, the Fourier
coefficients of the function $h$, as a result  of  the truncation of  the 
series. The contribution of  terms higher than $N$ in the expansion,
$\Delta_{N} h_{n_1 n_2}(\epsilon)$, can be 
estimated in terms of the function $\varphi(k)$ and Eq.~\ref{HCOEFS}.
The appropriate formula is 
\be
\label{ERROR_H}
(-1)^{n_1+n_2}\,  \Delta_{51} h_{n_1 n_2}(\epsilon)  = -\frac{1}{\xi}
\int_{51/\xi}^\infty dy\, \varphi(y)\, \epsilon^{\xi y}\approx
\frac{Z_1(e^{-1/\xi} \epsilon)^{51} }{2\xi(1-\xi \ln{\epsilon})}
\ee
The last equality is obtained from the asymptotic behaviour of 
$\varphi(y)$ and, thus, is only valid for $(n_1^2+n_2^2)<10$.  
Applying this formula we get results which match with the discrepancies 
observed in some cases. For example, as we said before,  $Y(0)$ ($\epsilon=1$)
should be equal to $0.5$ irrespective of which mode is used to compute it. 
However, our formula Eq.~\ref{ERROR_H} predicts that the truncated evaluation 
up to 51 orders and $n_1=n_2=2$ should fall short by $0.1105$.
The actual discrepancy found previously  is $0.1122$. Everything 
fits nicely. Our formulas can also be used to estimate the number of terms 
in the expansion required to obtain a given Fourier coefficient on the 
torus with a certain precision.

\subsection{$q=2$}

Here we will apply our method to  a multivortex situation. We take unit
aspect ratio ($\tau=1$) and two units of flux. We can also use the 
procedure explained previously to fix the position of the zeroes of the 
Higgs field. We took the following points:
\begin{equation}
  \left(0.35l_1, \frac{l_2}{2} \right); 
  \left(0.65l_1, \frac{l_2}{2} \right);
\end{equation}
which are separated along the $x$ direction a distance $0.3 l_1$. 

In  Figs.~7,8 we display the shape of the magnetic field
obtained for $\epsilon=0.9$ and 51 orders in the expansion. There is no
particular difference in computational cost between this case and the unit 
flux one. The effects of truncation are similar to those
obtained in the previous section:  modes up to 
$\max(n_1,n_2)<15$ are calculated up to machine precision. 
Note, however, that the position of the zeroes introduces a new scale 
in the problem, which translates into a typical scale for the modes.
This might cause trouble if the zeroes are very close
together.

We have repeated all our previous analysis of convergence
with qualitatively identical results. For example, the $L_2$ norm of the 
function $(B(x)-\frac{1}{2}(1-|\phi|^2))$, noted 
$L_2(N,\epsilon)$, seems to fall off exponentially fast with $N$, as in
the  $q=1$ case. With similar definitions and methods to the ones used 
for $q=1$ we got $a(\epsilon)$ and $b(\epsilon)$. Our best fit to the 
former quantity now gives:
\begin{equation}
  a(\epsilon) = 1.032(6)\log_{10}\epsilon - 0.020(1)
  \end{equation}
Our previous conclusions about convergence extend to this case as well.

Focus
It is interesting to compare the precision of our results with those 
employing other methods. In particular, the shape of the solution 
was computed

\DOUBLEFIGURE[h]{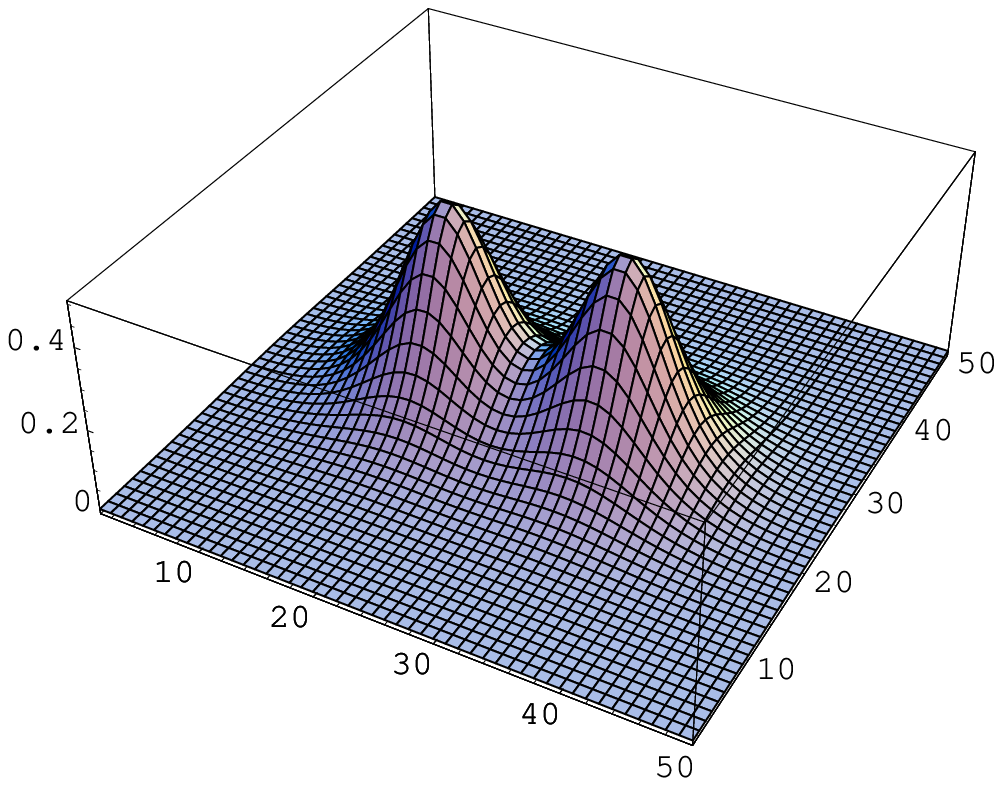,width=6cm}{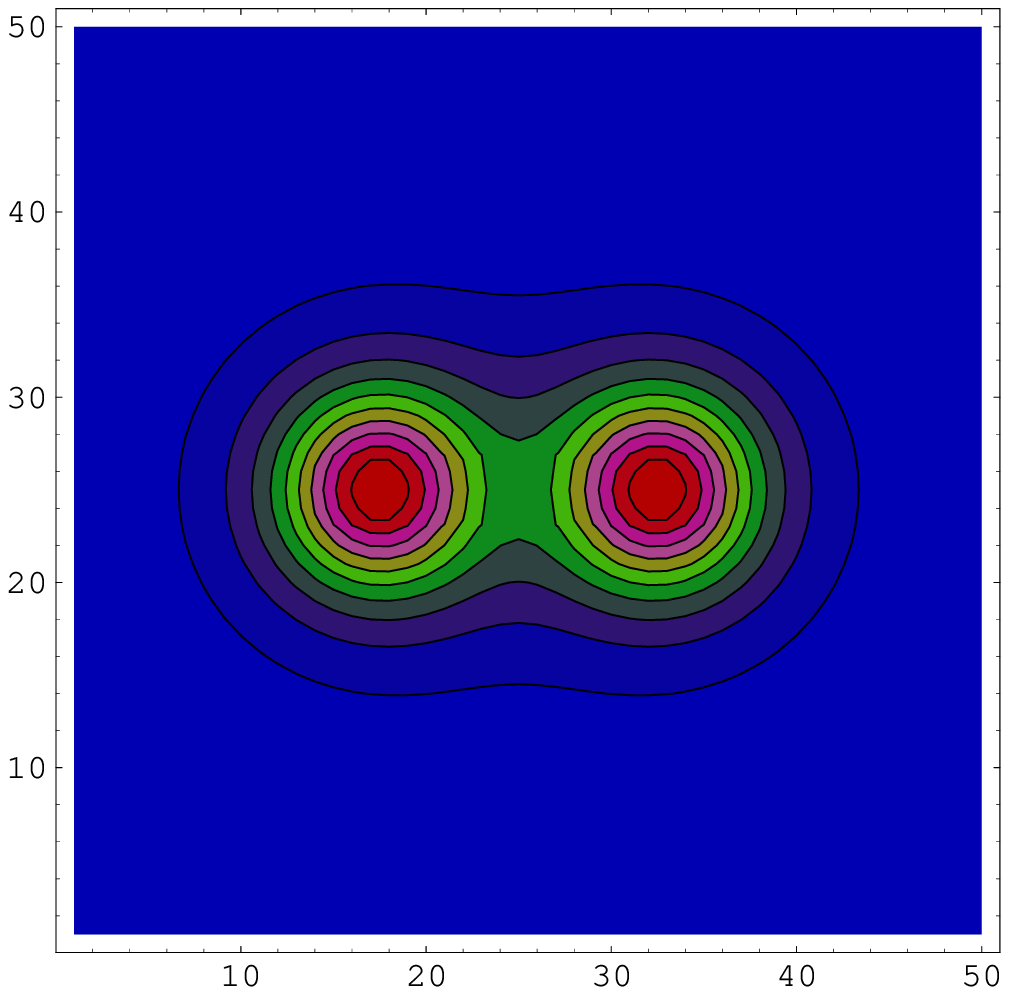,width=6cm}{$B(x)$ field for $\epsilon=0.9$, and 51 orders.}{Contour of $B(x)$ for $\epsilon=0.9$, and 51 orders.}

Focusing on the results for  large areas, we emphasise that 
the main effect of a change in 
$\epsilon$ will the be  to  vary the separation between the vortices. 
Thus, with the coefficients obtained from our analysis we can actually 
explore the variations in shape  for nearby vortices as a function 
of separation, a study which can 
have some interest (see for example Ref.\cite{Burzlaff:2000wv}). 
In this ($\epsilon \sim 1$) case, we can also compare with other alternative 
methods. In particular,  in Ref.~\cite{Samols:ne} the two vortex solution is 
computed by numerical methods. The finite square size used corresponds to 
$\epsilon=0.91$ and the precision attained $10^{-4}$. Our formulas give 
a precision in the $10^{-5}-10^{-4}$ for this size, which is, at least,
as good.

\section{Conclusions}

Let us summarise our results. We have shown how one can expand the 
solutions of the Bogomolny equations on a two dimensional torus 
in powers of $\epsilon=1-2f$, where $f$ is the average magnetic 
field (flux over area). The coefficients of the Fourier modes can be
constructed using an iterative procedure involving convolutions. 
Although, no close analytical expression for the coefficients exist 
beyond the first non-trivial order, these coefficients 
can be determined up to double precision machine accuracy 
(15-17 significant digits), by truncating to a finite number of modes. 
This method can be applied to construct solutions with arbitrary flux 
and location of the Higgs field zeroes. We have computed the
coefficients for a couple of cases ($q=1$ and $q=2$) and the results
are very encouraging. The 51 order truncated expansion  
is estimated to describe the shape of the function within machine precision
up to areas which are two and a half times the critical area. 
But meaningful values can be extracted also 
for large sizes, where due to the exponential localisation of the solution, 
the configurations are close to those of infinite area. In particular, these 
results match nicely with what is known about the unit-vortex 
on the plane. Turning the information around, this allows to obtain precise 
expectations about the behaviour of the coefficients for large order, which 
are satisfied by the data. No significant difference in performance is
observed when studying multivortex solutions. 

The solutions on the torus are relevant to depict the behaviour of the 
system in a situation of high vortex density. Their thermodynamics
was analysed in Ref.~\cite{Shah:1993us}.
The description in terms of Fourier modes has been used previously in 
other contexts, like in the study of skyrmion 
crystals Ref.~\cite{kugler,castillejo}. It seems, however,  on the basis
of our results, that our method can be used successfully to study 
the infinite area case as well.

 There are a number of possible applications and generalisations of the
method to other problems or situations, some of which are currently under 
study. Special mention deserves the application to self-dual configurations
on the four dimensional torus. In this case, there are no explicit 
analytic formulas for the solutions and the present method might provide 
good results, as an alternative to numerical methods~\cite{GarciaPerez:1989gt}.
Actually, the first term in the expansion was obtained
previously~\cite{GarciaPerez:2000yt} and served as initial motivation of this
work. 

There are other interesting problems for which the present method can be
used. For example, in  the study of  the dynamics of vortices, specially
in the low energy limit in which the geodesic approximation is
valid~\cite{Manton:1981mp,Stuart:tc}. 
The main issue here is  the determination of the metric within the manifold 
of solutions. This metric can be extracted from the behaviour of the solution 
itself in the vicinity of the Higgs zeroes~\cite{Samols:ne}, which suggests that 
our method can be successfully used\cite{in_prep}.  This study offers the 
opportunity to express and analyse the conserved quantities  studied in
Ref.~\cite{mantonnasir} within our formalism.

Finally, it is interesting to notice that the critical area case 
has a generalisation   to Higgs-gauge systems in any Kahler manifold, 
in  what is called the Bradlow limit\cite{Bradlow:ir}. From this point of 
view, our expansion can be viewed as a particular case of a much more
general concept: an expansion in the Bradlow parameter. Although part 
of our technology is specific to the torus, we think that  there are 
appropriate generalisations  to other manifolds and Riemannian metrics 
by using a different set of basis functions. Indeed, the lowest term 
has already studied for  the case of the
two-sphere\cite{Baptista:2002kb}. These generalisations are  currently 
under study by the present authors. 

\appendix 

\section{Quantum mechanical formalism}
\label{sc:quantum}

In this section we will describe the formalism in quantum mechanical terms.
Here we follow the spirit, notation and formulas  of Ref.~\cite{Giusti:2001ta}.
Our goal is to characterise  the space of sections of a U(1) vector bundle on the 
two-dimensional torus  within a fixed trivialisation. 

Fields satisfying the boundary conditions (\ref{BCondA}-\ref{BConds}) make up
a pre-Hilbert space ${\cal H}_q$ with scalar product
\be
\label{eq:scalarprod}
\langle\Phi|\Psi\rangle= \frac{1}{\cal A} \int_{T^2}dx\; \Phi^*(x)\Psi(x)
\ee 
where ${\cal A}$ is the area of the torus. 
The integration is over the torus and because of the
integrand periodicity can be performed over any  fundamental cell.

Following the standard quantum mechanical procedure we will now look for 
a complete set of commuting operators which can serve to find a basis 
of the Hilbert space. One family of operators ${\mathbf U}_{\tilde{a}}$ is labelled by elements
$\tilde{a}$ of the dual lattice $\Lambda^*$:
\be
{\mathbf U}_{\tilde{a}}\Psi(x)= \exp\{2 \pi \imath \tilde{a}(x)\}  \Psi(x)
\ee
where $\tilde{a}$ is a linear function of $x$ satisfying $\tilde{a}(e^{(i)})\in {\mathbf Z}$.
Notice that we can introduce two special elements $\omega_i$ of $\Lambda^*$
associated with the form $\omega$:
\be
\omega_i(x)=\frac{\omega(e^{(i)},x)}{q}
\ee
The set of these two  elements is  the dual basis to $\{e^{(i)}\}$. 
All the  ${\mathbf U}_{\tilde{a}}$ operators are mutually commuting. 
They are just gauge transformations of a special kind. 

In addition we will look at operators implementing translations. However,
ordinary translations map out of ${\cal H}_q$ because the transition functions
depend on $x$. It is clear on the basis of the nature of our fields that we
should replace translations by appropriate parallel transporters. For 
that purpose we will make use of the privileged   connection $A^{(0)}$ 
on the torus, having constant (or uniform) field strength:
\be
A^{(0)}= \pi \omega(x,dx)
\ee

Finite translation operators ${\cal T}_a$ can be defined as parallel transporters
along straight lines:
\be
({\cal T}_a\Psi)=e^{-\imath \int_\gamma A^{(0)}}\; \Psi(x+a)= e^{-\imath
\pi\omega(x,a)}\,  \Psi(x+a)
\ee
Obviously these operators do not commute: their commutator is
determined by the flux of the gauge field through the corresponding
parallelogram.

Our boundary conditions can be then reformulated by saying that elements 
of ${\cal H}_q$, are those fields left invariant by the operators:
\be
{\cal T}_{e^{(i)}} {\mathbf U}_{(-q w_i)}
\ee
We can then introduce the operators
\be
K^{(i)}= {\cal T}_{-e^{(i)}/q}\, {\mathbf U}_{ w_i}
\ee
and re-express the boundary conditions by
saying that $(K^{(i)})^q={\mathbf I}$. Furthermore the operators obey:
\be
K^{(1)} K^{(2)}= \exp\{ 2 \pi \imath/q   \}\  K^{(2)} K^{(1)}
\ee
The operator $K^{(1)}$ generates a $Z_q$ group
and its eigenvalues are given by $\exp\{-2 \pi \imath\frac{s}{q}\}$,
where $s$ is an integer modulo $q$. Thus, one can decompose 
${\cal H}_q$ into the $q$ orthogonal subspaces of eigenvectors:
\be
 {\cal H}_q = \oplus_{s=1}^q  {\cal H}_{q, s}
\ee
$K^{(2)}$ maps ${\cal H}_{q, s}$ into ${\cal H}_{q, s-1}$.
The translation operators ${\cal T}_a$ commute with 
$K^{(i)}$ and hence leave these subspaces  invariant.

Our task of finding a complete set of operators is achieved by
$K^{(1)}$ and an operator involving translations. As we will see in the 
study of the Bogomolny equation it is natural to select this operator to be 
the covariant Laplacian constructed with the   constant field strength gauge
potential. Up to now the choice of metric in space has played no basic role. 
Here, however, this operator depends on the  metric. We will take the 
metric to be Euclidean. Within this metric the lattice vectors $e^{(i)}$ have 
a well defined length and scalar product. It is always possible to make a
coordinate transformation to bring $e^{(1)}$ to the form $(l_1,0)$. The other 
vector $e^{(2)}$ is then given by $l_2(\cos\varphi,\sin\varphi)$. This
reduces for $\varphi=\frac{\pi}{2}$ to the special case considered in
section~\ref{sc:method}.

The generators of translations along each axis are precisely the components 
of the covariant derivative corresponding to the $A^{(0)}$ field.
We obtain
\be
D^{(0)}_i=\partial_i + \pi \imath q \epsilon_{ i j} x_j /{\cal A}
\ee
where $\epsilon_{ i j}$ is the antisymmetric tensor with two 
indices($\epsilon_{1 2}=1$), and ${\cal A}$ is the area of the fundamental
cell. These operators are anti-hermitian and satisfy the following
commutation relations:
\be
\label{HAlgb}
[D_1^{(0)},D_2^{(0)}]=-\imath f\equiv\frac{ -2 \pi \imath q}{{\cal A}} 
\ee
After an appropriate rescaling this is just the Heisenberg algebra
satisfied by momentum and position operators in one dimensional 
Quantum Mechanics. Using standard formulas one can construct operators 
$\aa$, $\aa^\dagger$ satisfying the commutation relations of 
creation-annihilation operators, and express the covariant derivatives 
in terms of them:
\bea 
D_1^{(0)}=i \sqrt{\frac{f}{2}}(\aa+\aa^\dagger) \\
D_2^{(0)}=\sqrt{\frac{f}{2}}(\aa-\aa^\dagger)
\eea
The covariant Laplacian associated to the $A^{(0)}$ field is proportional to 
the Hamiltonian of a harmonic oscillator:
\be
D_i^{(0)}D_i^{(0)}=-f(\aa^\dagger \aa+\frac{1}{2})
\ee
Thus the space of {\em classical} sections of a U(1) bundle, has identical 
structure as the Landau levels of the quantum system.   

Finally,  a basis of 
our space of sections is provided by the 
states  $|n,s\rangle$ which are simultaneous eigenstates of the number
operator and  $K^{(1)}$, where $n$ is an arbitrary non-negative integer and 
$s$ an integer modulo $q$. 
We have the following relations
\bea
K^{(1)}|n,s\rangle = e^{-i\frac{2\pi s }{q}} |n,s\rangle \\
K^{(2)}|n,s\rangle = |n,s-1\rangle \\
D_1^{(0)}|n,s\rangle = i \sqrt{\frac{f}{2}}\, (\sqrt{n+1} |n+1,s\rangle+
\sqrt{n} |n-1,s\rangle)\\
D_2^{(0)}|n,s\rangle =\sqrt{\frac{f}{2}}\,  (-\sqrt{n+1} |n+1,s\rangle+
\sqrt{n} |n-1,s\rangle)
\eea
We consider the $|n,s\rangle$ states to be orthonormal within the scalar product 
Eq.~\ref{eq:scalarprod}.

We will now give  the  explicit form of the functions $\Psi_{n,s}(x_1,x_2)$ 
corresponding to these states ($|n,s\rangle$). For that purpose notice that  any function
$\Psi(x_1,x_2) \in {\cal H}_q$ can be expressed as:
\be
\label{interm}
\Psi(x_1,x_2)= e^{-\imath \pi q\omega_1(x) \omega_2(x)}\, \sum_{p \in
{\mathbf Z}} e^{-2 \pi \imath p \omega_2(x)} {\cal J}_p(\omega_1(x))
\ee
which follows from analysing the periodicity under $e^{(1)}$. Now imposing
that $\Psi$ belongs to ${\cal H}_{q, s}$, one concludes that, in the 
sum  appearing in Eq.~\ref{interm}, $p$ is restricted to $p=s\bmod q$. 
Imposing now periodicity under $e^{(2)}$ we arrive at:
\be
\label{changebasis}
\Psi(x_1,x_2)= 
e^{-\imath \pi q\omega_1(x) \omega_2(x)}\, \sum_{p \in
s+q{\mathbf Z}}
 e^{-2 \pi \imath p \omega_2(x)}\,  {\cal J}(q\omega_1(x)+p)
\ee
Now we look at eigenstates of the number operator. For that purpose it is
interesting to look at the way in which the creation and annihilation 
operators act on the function ${\cal J}$ appearing in the previous formula. 
We have:
\bea
(\aa {\cal J})(y)=\frac{1}{\sqrt{2}} (
\frac{d}{dy'}-\imath\frac{e^{i\varphi}}{\sin\varphi} y'){\cal J}(y)\\
(\aa^\dagger {\cal J})(y)=\frac{-1}{\sqrt{2}} (
\frac{d}{dy'}-\imath\frac{e^{-i\varphi}}{\sin\varphi} y'){\cal J}(y)
\eea
where $y'=\sqrt{\frac{2 \pi l_2 \sin\varphi}{q l_1}} y$.

After a standard quantum mechanical calculation we arrive at 
the expression of the function  $\Psi_{n s}(x_1,x_2)$
corresponding to the state 
$|n,s\rangle$
\be
\Psi_{n s}(x_1,x_2)=(\frac{2 q l_2\sin\varphi}{l_1})^{1/4}\, e^{-\imath \pi q\omega_1(x) \omega_2(x)}\, \sum_{p \in
s+q{\mathbf Z}} e^{-2 \pi \imath p \omega_2(x)} e^{\imath\frac{e^{\imath
\varphi}}{2 \sin\varphi}y^{\prime 2}} H_n(y')
\ee
where $H_n$ is a Hermite polynomial and $y'=\sqrt{f}(x_2+\frac{p l_2
\sin\varphi}{q})$. To reduce to the case of orthogonal $e^{(i)}$ one 
can fix $\varphi=\frac{\pi}{2}$, $\omega_1(x)=x_2/l_2$ and $\omega_2(x)=-x_1/l_1$.

For the special case of the vacuum state ($n=0$), we get  
\be
\Psi_{0 s}(x_1,x_2)=(\frac{2 q l_2\sin\varphi}{l_1})^{1/4}\, e^{-\imath \pi q\omega_1(x)\omega_2(x)}\, 
e^{ \imath \pi \tau \omega_1^2}\,  \thetachar{s/q}{0}(z,\tau)
\ee
where  $\thetachar{s/q}{0}(z,\tau)$  is a theta function 
with rational characteristics with  arguments:
\bea
z=\pi(\tau \omega_1-q \omega_2)  \\
 \tau= e^{\imath
 \varphi} q \frac{l_2}{l_1}
\eea

Now we need to deduce the
action of the translation operator ${\cal T}(a)$ on these states. 
The translation operator is defined as  ${\cal
T}(a)=\exp\{a_1D_1^{(0)}+a_2D_2^{(0)}\}=\exp\{-(z(a) \aa
 -z^*\aa^\dagger )\}$. 
Now we can compute the matrix elements 
\bea
\langle m,s'|{\cal T}(a)|n,s\rangle = \delta_{s s'}\, 
e^{i \beta(n-m)} (-1)^{(M+n)}\times \\ 
\nonumber \,
e^{-\frac{|z|^2}{2}} |z|^{|m-n|} \sum_{j=0}^{M} (-1)^j
\frac{\sqrt{n!m!}\, |z|^{2j}}{j! (M-j)! (j+|m-n|)!}
\eea
where $M=\min{(m,n)}$ and 
\be
\label{ZFORM}
z(a)=-i\sqrt{\frac{f}{2}}(a_1-ia_2)\equiv |z| e^{i \beta}
\ee

We can also compute  the matrix elements of the operators ${\mathbf U}_{\tilde{a}}$.
Expanding $\tilde{a}$ in the dual basis  of $\Lambda^*$ we can write $\tilde{a}(x)=-k_1
\omega_2(x)+k_2
\omega_1(x)$. We will then denote  ${\mathbf U}_{\tilde{a}}\equiv
U(k_1,k_2)$. This operator can be expressed  in terms of $K^{(i)}$ and translations as follows: 
\be
U(k_1,k_2)=e^{-i\pi\frac{k_1k_2}{q}}\, (K^{(2)})^{-k_1}  (K^{(1)})^{k_2}  
{\cal T}(a(k_1,k_2))
\ee
From here we can compute the matrix elements:
\bea
\langle m,s'| U(k_1,k_2) |n,s\rangle= \delta_{s'\, s+k_1} e^{2 \pi i \frac{s
k_2}{q}}\, e^{-i\pi\frac{k_1k_2}{q}}\, e^{i \alpha(n-m)} (-1)^{(M+n)}\times \\ 
\nonumber \,
e^{-\frac{\tilde{\xxi}}{2}} \tilde{\xxi}^{|m-n|/2} \sum_{j=0}^{M} (-1)^j
\frac{\sqrt{n!m!}\, \tilde{\xxi}^{j}}{j! (M-j)! (j+|m-n|)!}
\eea
where $\alpha$ and $\tilde{\xxi}$ can be obtained from the complex number
\be
\ttau=-\imath
\sqrt{\frac{f}{2}}(l_1k_2-l_2k_1e^{i\varphi})/q=\sqrt{\tilde{\xxi}}\,e^{\imath
\alpha}
\ee
and $\tilde{\xxi}$ is the generalisation of $\xxi$ defined in
Eq.~\ref{eq:defxi} to the case of non-orthogonal $e^{(i)}$:
\be
\tilde{\xxi} = |\ttau|^2= \frac{\pi}{q  \sin\varphi}( k_1^2 \frac{l_2}{l_1} + k_2^2
\frac{l_1}{l_2
}-2 k_1k_2 \cos\varphi)
\ee

\subsection{The Bogomolny equations}

We can now reformulate  the Bogomolny equations in this formalism. We 
can write the Higgs field as $\phi=\epsilon\psi$, where $\psi$ 
is a normalised element of ${\cal H}_q$. It  can be decomposed 
in our basis  
\be
\psi=\sum_{n=0}^{\infty}c(n,s)\, |n,s\rangle
\ee
where the sum of the modulus square of the coefficients $c(n,s)$ equals unity.
Similarly the function $h$ appearing in Section~\ref{sc:method} is expressed in 
terms of a generalised Fourier decomposition
\be
h(x_1,x_2)=\sum_{k_1 k_2}\, h_{k_1k_2} e^{2 \pi i (-k_1
\omega_2(x)+k_2
\omega_1(x))}\,
\ee

The first Bogomolny equation can then be expressed as follows:
\be
\aa\psi=i \frac{1}{\sqrt{2f}}\left( (\partial_1+i\partial_2)h \right)\psi
\ee
Hence, using the decompositions of $h$ and $\psi$ and the matrix elements
deduced in this Section, we can re-express the Bogomolny equations as follows:
\bea
\label{BOGNONE}
(1-\epsilon)|\ttau|^2 h_{k_1k_2}=\frac{\epsilon}{2}\sum_{m,s',n,s}
\, c^*(m,s')\, c(n,s)\,  \langle m, s'|U(-k_1, -k_2)|n, s\rangle\\
\label{BOGNTWO}
\sqrt{n+1}\, c(n+1,s')=-\sum_{k_1 k_2}\, \ttau^* h_{k_1 k_2}\, \sum_{p, s}\, c(p, s)\,
\langle n, s'|U(k_1, k_2)|p, s\rangle  
\eea
where $\ttau=\imath \sqrt{\frac{f}{2}}(k_1l_2e^{-\imath \varphi}-k_2l_1)/q$. 
The volume dependence of these equations is explicit (contained in the
dependence on $\epsilon$). Our method consists then in expanding 
the unknown coefficients in powers of  $\epsilon$:
\bea
c(p, s)= \sum_{k=0}^{\infty} c^{(k)}(p, s)\, \epsilon^k\\
h_{k_1k_2}= \sum_{k=1}^{\infty} h^{(k)}_{k_1k_2}\, \epsilon^k 
\eea
and solving the equations order by order. 

Let us compute the leading terms for the $q=1$ case. To lowest non-trivial
order we have 
\be
c^{(0)}(n)=\delta_{n 0}
\ee
and $h^{(0)}_{k_1k_2}=0$. 
Then, plugging this value into the first Bogomolny equation we get:
\be
h^{(1)}_{k_1k_2}=\frac{1}{\tilde{\xxi}}\langle 0|U(-k_1
-k_2)|0\rangle=\frac{(-1)^{k_1k_2}}{2\tilde{\xxi}}\,
e^{-\tilde{\xxi}/2} 
\ee
The following order obtains  from the other equation
\be
c^{(1)}(n)=-\frac{1}{\sqrt{n}} \sum_{k_1 k_2}\, \ttau^* h^{(1)}(k_1, k_2)\, 
\langle n-1|U(k_1, k_2)|0\rangle= -\frac{1}{2\sqrt{n!}}  \sum_{k_1 k_2}\,
\frac{(\ttau^*)^{n}}{\tilde{\xxi}}\, e^{-\tilde{\xxi}}
\ee
This is real and  vanishes for odd  $n$. 

One can then continue to iterate Eqs.~\ref{BOGNONE}--\ref{BOGNTWO} to higher orders
in $\epsilon$. To evaluate the expressions one has to restrict to a finite
number of elements on the basis. Now, in  addition to the cut in Fourier modes
$n_{\mbox{\tiny max}}$, one has to cut in the spectrum of the number operator 
$n\le n_{\mbox{\tiny part}}$. As the order $k$ grows  both $n_{\mbox{\tiny max}}$
and $n_{\mbox{\tiny part}}$ should grow to keep the numbers within machine
precision. We have developed an independent Fortran90  program to evaluate 
the $c^{(k)}(n)$, $h^{(k)}(n_1,n_2)$ using this procedure. 
With this program we have reproduced, within machine
precision, the coefficients in the expansion of the low lying modes $n_i$ up
to 30th order in the expansion for the $q=1$ case.  This serves as a check
that there are no unexpected bugs in the determination of the coefficients.
Increasing the order one starts noticing sizable errors associated to the 
cut in $n_{\mbox{\tiny part}}$. Unfortunately, increasing this number is
not only limited by computer resources, but also by numerical instabilities. 
For example the computation of the matrix elements
$\langle n, s'|U(k_1, k_2)|p,s\rangle$ becomes unstable for large $k_i$, $n$
and $p$. This is due to wild cancellations of large numbers appearing in
their definition. Actually, the problem already appears at lower orders.
In our calculation to order 30, we  had to tabulate
these matrix elements and compute them independently with a
\texttt{C} code and the GNU Multiple Precision Arithmetic Library
(\texttt{GMP}) which allows arbitrary precision floating point
operations. The values of the 
matrix elements were computed up to $n_{\mbox{\tiny part}}=150$
and performing intermediate calculations with $300$ significant decimal digits.
In summary, it turns out that this method is less efficient than
the one explained in section~\ref{sc:method}. Nevertheless, apart from serving as a
check of the results, we found interesting to explore this {\em quantum
mechanical} procedure, since it seems directly generalisable to the 
construction of self-dual configurations in the four-dimensional
torus~\cite{GarciaPerez:2000yt}, and all the main intermediate objects 
(as $\langle n, s'|U(k_1, k_2)|p,s\rangle$) appear there as well. 
This is currently under study.

\acknowledgments

The authors acknowledge financial support from the Spanish Ministerio de C\'{\i}encia 
y Tecnolog\'{\i}a under grant FPA2003-03801.

\end{document}